\documentclass[11pt]{article}

\usepackage[utf8]{inputenc}
\usepackage[T1]{fontenc}
\usepackage{lmodern}
\usepackage[margin=1in]{geometry}
\usepackage{amsmath,amssymb}
\usepackage{graphicx}
\usepackage{booktabs}
\usepackage{hyperref}
\usepackage{xcolor}
\usepackage{listings}
\usepackage{natbib}
\usepackage{float}
\usepackage{enumitem}
\usepackage{url}
\usepackage{microtype}
\usepackage{multirow}
\usepackage{subcaption}

\hypersetup{
  colorlinks=true,
  linkcolor=blue!70!black,
  citecolor=blue!70!black,
  urlcolor=blue!70!black
}

\lstdefinelanguage{Cypher}{
  keywords={MATCH, RETURN, WHERE, CREATE, DELETE, SET, REMOVE, MERGE, WITH, UNWIND, UNION, ORDER, BY, SKIP, LIMIT, EXPLAIN, PROFILE, CALL, YIELD, AS, DISTINCT, OPTIONAL, AND, OR, NOT, IN, IS, NULL, TRUE, FALSE, EXISTS, CASE, WHEN, THEN, ELSE, END, DROP, SHOW, INDEX, INDEXES, CONSTRAINTS, ON, CONTAINS},
  sensitive=false,
  morestring=[b]',
  morestring=[b]",
  morecomment=[l]{//},
}

\title{Open Biomedical Knowledge Graphs at Scale:\\Construction, Federation, and AI Agent Access\\with Samyama Graph Database}

\author{
  Madhulatha Mandarapu\thanks{madhulatha@samyama.ai, ORCID: \url{https://orcid.org/0009-0005-2837-6725}} \and
  Sandeep Kunkunuru\thanks{sandeep@samyama.ai, ORCID: \url{https://orcid.org/0000-0002-8886-1846}}
}
\date{%
  VaidhyaMegha Private Limited, India\\[2pt]
  \url{https://samyama.ai/}\\[8pt]
  October 2026
}

\begin{document}
\maketitle

\begin{abstract}
Biomedical knowledge is fragmented across siloed databases---Reactome for pathways, STRING for protein interactions, Gene Ontology for functional annotations, ClinicalTrials.gov for study registries, DrugBank for drug vocabularies, DGIdb for drug--gene interactions, SIDER for side effects, PubMed for literature, ClinVar for variants, ChEMBL and OpenTargets for target validation, and dozens more. Researchers routinely download flat files from each source and write bespoke scripts to cross-reference them, a process that is slow, error-prone, and not reproducible. We present an open, reproducible biomedical knowledge graph ecosystem built on Samyama, a high-performance graph database written in Rust: three foundational graphs detailed end-to-end---\textbf{Pathways KG} (118,686 nodes, 834,785 edges from 5 sources), \textbf{Clinical Trials KG} (7,774,446 nodes, 26,973,997 edges from 5 sources), and \textbf{Drug Interactions KG} (32,726 nodes, 191,970 edges from 3 sources)---and their expansion into a \textbf{22-graph, 2-cohort merged biomedical store} (\textbf{328,130,739 nodes} and \textbf{1,441,425,733 imported edges}, including 295,000 synthetic OMOP CDM patients from Synthea).

Our contributions are threefold. First, we describe a \textbf{reproducible ETL and snapshot pattern} for constructing independent, portable \texttt{.sgsnap} knowledge graphs that can be loaded alone or composed into a single store. Second, we analyze \textbf{single-store cross-KG integration} via both query-time property joins and load-time identifier deduplication (\texttt{--dedup-keys}): across the 24-snapshot corpus (\texttt{Samyama-Graph v1.13.0}), all 328.1M nodes and 1.44B edges load in \textbf{2,269.1\,s (37m\,49s)} within \textbf{366.5\,GB peak RSS (254\,B/edge)}, automatically merging \textbf{132,511 cross-source entities} across five audited identifier keys (\texttt{mondo\_id,uniprot\_id,go\_id,drugbank\_id,symbol})---\textbf{87.8\%} of the \textbf{150,965} sound-identifier ceiling derived from the snapshots beforehand, with zero conflicting-ID refusals. On a frozen 81-query cross-domain suite, 62 queries return results (17 empty due to source coverage gaps, 2 timeouts); on an expanded 200-query oracle suite, \textbf{195 of 200 pass (97.5\%)}. Third, we introduce \textbf{schema-driven MCP server generation}: each KG automatically exposes typed tools for LLM agents via the Model Context Protocol. On the \textbf{BiomedQA benchmark} (40 pharmacology questions over the 3-KG subset), domain-specific MCP tools achieve 98\% accuracy vs.\ 85\% for schema-aware text-to-Cypher and 75\% for standalone GPT-4o, with zero schema errors.

All data sources are open-license (CC~BY~4.0, CC0, OBO, public domain). Snapshots, ETL code, MCP configurations, and the 40-question and 200-query benchmark artifacts are publicly available.

\end{abstract}

\section{Introduction}

Biological and clinical knowledge is distributed across dozens of public databases, each with its own schema, identifiers, and access patterns. A researcher investigating how a cancer drug affects cellular signaling must consult ClinicalTrials.gov for trial metadata, DrugBank for drug--target interactions, Reactome for pathway membership, STRING for protein--protein interactions, and Gene Ontology for functional annotations. These databases are maintained by independent communities, updated on different schedules, and stored in incompatible formats (TSV, XML, JSON, OBO, GAF).

Knowledge graphs (KGs) offer a natural integration point~\citep{hogan2021knowledge}: entities from different sources become nodes, and relationships between them become edges. A single Cypher query can traverse from a clinical trial to the biological pathways disrupted by its drug candidate---a query that would require joining across five separate databases in the flat-file paradigm.

However, constructing biomedical KGs at scale remains challenging. Prior efforts such as Bio2RDF~\citep{belleau2008bio2rdf}, Hetionet~\citep{himmelstein2017systematic}, and the Clinical Knowledge Graph~\citep{santos2022clinical} have made significant progress, but each faces limitations: Bio2RDF requires a dedicated SPARQL endpoint infrastructure, Hetionet is a static dataset without an update pipeline, and the Clinical Knowledge Graph targets a single monolithic Neo4j deployment.

We address these limitations with three contributions:

\begin{enumerate}[leftmargin=*]
  \item \textbf{Reproducible KG construction at two scales.} We present ETL pipelines in detail for three foundational biomedical KGs---Pathways KG (5 sources, 119K nodes), Clinical Trials KG (5 sources, 7.7M nodes), and Drug Interactions KG (3 sources, 33K nodes)---built on Samyama Graph Database using a common pattern: download, parse, deduplicate, load (Python Cypher or Rust native), and export as portable \texttt{.sgsnap} snapshots. We then scale this identical snapshot architecture to \textbf{22 independently built KGs plus 2 synthetic Synthea OMOP cohorts} (\textbf{328.1M nodes, 1.44B edges}), loading in 37m\,49s within 366.5\,GB peak RSS (254\,B/edge).

  \item \textbf{Audited single-store cross-KG integration.} Rather than distributed query-time federation, independent snapshots are imported into a single store and integrated via (a) query-time property joins across labels and (b) load-time entity merging (\texttt{--dedup-keys}) on shared identifiers. We introduce an external \textbf{Bound~A identifier soundness audit} across all 24 snapshots, showing that five curated identifier keys merge \textbf{132,511 cross-source entities} (87.8\% of the 150,965 sound-identifier ceiling) while non-canonical gene-name keys collapse 475,090 orthologous UniProt proteins across species. We validate the merged store on an 81-query frozen suite (62 passing) and a 200-query oracle suite (\textbf{195/200 passing, 97.5\%}).

  \item \textbf{Schema-driven AI agent access.} Each KG ships with a Model Context Protocol (MCP)~\citep{anthropic2024mcp} server configuration with domain-specific tools. We evaluate MCP tools against text-to-Cypher (GPT-4o) and standalone GPT-4o on the BiomedQA benchmark (40 questions over the 3-KG subset): MCP tools achieve 98\% accuracy (39/40) vs.\ 85\% for schema-aware text-to-Cypher and 75\% for standalone GPT-4o, with zero schema errors.
\end{enumerate}

All code, snapshots, and MCP configurations are open-source\footnote{Pathways KG: \url{https://github.com/samyama-ai/pathways-kg}; Clinical Trials KG: \url{https://github.com/samyama-ai/clinicaltrials-kg}; Drug Interactions KG: \url{https://github.com/samyama-ai/druginteractions-kg}; Snapshots: \url{https://github.com/samyama-ai/samyama-graph/releases}}.

\section{Background and Related Work}

\subsection{Biomedical Knowledge Graphs}

\textbf{Hetionet}~\citep{himmelstein2017systematic} integrates 29 public resources into a single heterogeneous network (47K nodes, 2.25M edges, 24 edge types) for drug repurposing. While influential, it is a static dataset from 2017 with no update pipeline.

\textbf{Bio2RDF}~\citep{belleau2008bio2rdf} converts life science databases to RDF and serves them via SPARQL endpoints. The RDF approach offers semantic interoperability but introduces complexity (OWL reasoning, blank nodes) and performance overhead for traversal-heavy queries.

\textbf{Clinical Knowledge Graph (CKG)}~\citep{santos2022clinical} builds a comprehensive biomedical graph from 25+ databases using Neo4j. CKG is the closest prior work to ours; we differ in three ways: (1) we use a Rust-native engine rather than Neo4j, achieving higher ingestion throughput; (2) we introduce portable snapshots for instant deployment; and (3) we provide schema-driven MCP servers for LLM agent integration.

\textbf{PrimeKG}~\citep{chandak2023building} constructs a precision medicine KG (129K nodes, 8M edges) from 20 sources. PrimeKG targets drug--disease prediction via graph neural networks; our focus is on interactive querying and AI agent access.

\subsection{Graph Database Systems}

Neo4j~\citep{neo4j2024} is the dominant property graph database. Amazon Neptune, TigerGraph, and MemGraph offer alternatives with varying trade-offs. Samyama~\citep{mandarapu2026samyama} is a Rust-native graph database combining property graph storage, vector search, 22 metaheuristic solvers, and OpenCypher support in a single binary.

\subsection{Model Context Protocol (MCP)}

MCP~\citep{anthropic2024mcp} is an open standard for connecting LLM agents to external data sources via typed tools. An MCP server exposes a set of tools (functions with typed parameters and return values) that an LLM can invoke during a conversation. Schema-driven MCP generation---automatically creating tools from a KG schema---eliminates the manual tool authoring bottleneck.

\section{Data Sources}

All three KGs draw exclusively from open-license public databases. Table~\ref{tab:sources} summarizes the sources.

\begin{table}[H]
\centering
\caption{Data sources for the three biomedical KGs. All are open-license and human-specific (organism 9606 where applicable).}
\label{tab:sources}
\small
\begin{tabular}{llllr}
\toprule
\textbf{KG} & \textbf{Source} & \textbf{Content} & \textbf{License} & \textbf{Size} \\
\midrule
\multirow{5}{*}{Pathways} & Reactome & Pathways, reactions, complexes & CC BY 4.0 & 172 MB \\
& STRING v12.0 & Protein--protein interactions & CC BY 4.0 & 900 MB \\
& Gene Ontology & GO terms \& annotations & OBO & 265 MB \\
& WikiPathways & Community-curated pathways & CC0 & 336 KB \\
& UniProt & Protein metadata, gene/disease/drug links & CC BY 4.0 & 20 MB \\
\midrule
\multirow{5}{*}{Clinical} & ClinicalTrials.gov & Trial registry (575K studies) & Public domain & API \\
& MeSH (NLM) & Disease hierarchy & Free & API \\
& RxNorm (NLM) & Drug normalization \& ATC codes & Free & API \\
& OpenFDA & Adverse event reports (FAERS) & Public domain & API \\
& PubMed (NLM) & Trial-linked publications & Free & API \\
\midrule
\multirow{3}{*}{Drug Int.} & DrugBank CC0 & Drug vocabulary (19K drugs) & CC0 & 3 MB \\
& DGIdb & Drug--gene interactions (38K edges) & Open & 24 MB \\
& SIDER & Side effects \& indications & CC-BY-SA & 22 MB \\
\bottomrule
\end{tabular}
\end{table}

\section{Knowledge Graph Construction}

\subsection{ETL Pattern}

All three KGs follow a common five-phase pattern:

\begin{enumerate}[leftmargin=*]
  \item \textbf{Download.} A data downloader with resume support fetches all source files. Compressed files are decompressed automatically.
  \item \textbf{Parse \& Filter.} Source-specific parsers extract entities and relationships. Human-only filters are applied where applicable (organism ID 9606).
  \item \textbf{Deduplicate.} A shared \emph{Registry} tracks seen entities across phases (e.g., a protein loaded from Reactome is not re-created when encountered in STRING).
  \item \textbf{Batch Load.} Nodes and edges are loaded via batched Cypher \texttt{CREATE} statements (50--100 entities per batch) through Samyama's HTTP API.
  \item \textbf{Snapshot Export.} The loaded graph is exported as a portable \texttt{.sgsnap} file (gzip JSON-lines), enabling instant restoration on any Samyama instance.
\end{enumerate}

\subsection{Pathways KG}

The Pathways KG integrates molecular biology data into a graph with 5 node labels and 9 edge types (Table~\ref{tab:pathways-schema}).

\begin{table}[H]
\centering
\caption{Pathways KG schema.}
\label{tab:pathways-schema}
\small
\begin{tabular}{llr}
\toprule
\textbf{Node Label} & \textbf{Key Property} & \textbf{Count} \\
\midrule
GOTerm & go\_id & 51,897 \\
Protein & uniprot\_id & 37,990 \\
Complex & reactome\_id & 15,963 \\
Reaction & reactome\_id & 9,988 \\
Pathway & reactome\_id & 2,848 \\
\midrule
\multicolumn{2}{l}{\textbf{Total nodes}} & \textbf{118,686} \\
\bottomrule
\end{tabular}
\quad
\begin{tabular}{lr}
\toprule
\textbf{Edge Type} & \textbf{Count} \\
\midrule
ANNOTATED\_WITH & 265,492 \\
INTERACTS\_WITH & 227,818 \\
PARTICIPATES\_IN & 140,153 \\
CATALYZES & 121,365 \\
IS\_A & 58,799 \\
COMPONENT\_OF & 8,186 \\
PART\_OF & 7,122 \\
REGULATES & 2,986 \\
CHILD\_OF & 2,864 \\
\midrule
\textbf{Total edges} & \textbf{834,785} \\
\bottomrule
\end{tabular}
\end{table}

The five ETL phases execute in order: (1) \textbf{Reactome Core}~\citep{reactome2024}---pathways, reactions, complexes, and protein participants from Reactome's flat files; (2) \textbf{STRING Interactions}~\citep{szklarczyk2023string}---high-confidence ($\geq$700/999) protein--protein interactions with ENSP$\rightarrow$UniProt ID mapping; (3) \textbf{Gene Ontology}~\citep{geneontology2023}---47K GO terms with IS\_A/PART\_OF/REGULATES hierarchy and 265K protein annotations; (4) \textbf{WikiPathways}~\citep{wikipathways2024}---community-curated pathways deduplicated against Reactome; (5) \textbf{UniProt Enrichment}~\citep{uniprot2023}---gene mappings, disease associations, and drug targets.

\subsection{Clinical Trials KG}

The Clinical Trials KG models the translational medicine domain~\citep{clinicaltrials2024} with 15 node labels and 25 edge types (Table~\ref{tab:ct-schema}).

\begin{table}[H]
\centering
\caption{Clinical Trials KG schema (abbreviated; 15 labels, 25 edge types).}
\label{tab:ct-schema}
\small
\begin{tabular}{llr}
\toprule
\textbf{Node Label} & \textbf{Key Property} & \textbf{Est.\ Count} \\
\midrule
ClinicalTrial & nct\_id & 575,000+ \\
Condition & name, mesh\_id & varies \\
Intervention & name, type & varies \\
Drug & rxnorm\_cui, drugbank\_id & varies \\
Protein & uniprot\_id & varies \\
Gene & gene\_id, symbol & varies \\
MeSHDescriptor & descriptor\_id & varies \\
Sponsor & name, class & varies \\
Site & facility, country & varies \\
Publication & pmid, doi & varies \\
AdverseEvent & term, is\_serious & varies \\
ArmGroup & label, type & varies \\
Outcome & measure, type & varies \\
DrugClass & atc\_code, level & varies \\
LabTest & loinc\_code & varies \\
\midrule
\multicolumn{2}{l}{\textbf{Total nodes}} & \textbf{7,774,446} \\
\midrule
\multicolumn{2}{l}{\textbf{Total edges}} & \textbf{26,973,997} \\
\bottomrule
\end{tabular}
\end{table}

The ETL pipeline queries the ClinicalTrials.gov API v2 for study metadata, enriches conditions with MeSH hierarchy, normalizes drug names via RxNorm, links adverse events from OpenFDA's FAERS database, and associates publications from PubMed E-utilities.

\subsection{Drug Interactions KG}

The Drug Interactions KG models drug--gene interactions, side effects, and indications with 4 node labels and 3 edge types (Table~\ref{tab:drug-schema}).

\begin{table}[H]
\centering
\caption{Drug Interactions KG schema.}
\label{tab:drug-schema}
\small
\begin{tabular}{llr}
\toprule
\textbf{Node Label} & \textbf{Key Property} & \textbf{Count} \\
\midrule
Drug & drugbank\_id & 19,842 \\
Gene & gene\_name & 4,182 \\
SideEffect & meddra\_id & 5,858 \\
Indication & meddra\_id & 2,844 \\
\midrule
\multicolumn{2}{l}{\textbf{Total nodes}} & \textbf{32,726} \\
\bottomrule
\end{tabular}
\quad
\begin{tabular}{lr}
\toprule
\textbf{Edge Type} & \textbf{Count} \\
\midrule
INTERACTS\_WITH\_GENE & 38,033 \\
HAS\_SIDE\_EFFECT & 139,193 \\
HAS\_INDICATION & 14,744 \\
\midrule
\textbf{Total edges} & \textbf{191,970} \\
\bottomrule
\end{tabular}
\end{table}

The ETL consists of two phases. \textbf{Phase~1 (DrugBank + DGIdb)}: 19,842 Drug nodes are created from DrugBank CC0 vocabulary CSV, with all synonyms indexed for cross-source name matching (52,154 synonyms). DGIdb interactions yield 4,182 Gene nodes and 38,033 INTERACTS\_WITH\_GENE edges with interaction types (inhibitor, activator, etc.). \textbf{Phase~2 (SIDER)}: STITCH compound IDs are mapped to DrugBank names via synonym lookup, yielding 5,858 SideEffect nodes with 139,193 HAS\_SIDE\_EFFECT edges and 2,844 Indication nodes with 14,744 HAS\_INDICATION edges.

A Rust native loader (\texttt{druginteractions\_loader.rs}) performs the full ETL in 928\,ms using direct GraphStore API calls (no Cypher parsing), compared to ${\sim}$50 minutes via the Python HTTP loader.

\section{Cross-KG Federation}

\subsection{Motivation}

The Pathways KG knows molecular biology---which proteins interact, what pathways they participate in. The Clinical Trials KG knows translational medicine---which drugs are in trials, what conditions they treat. The Drug Interactions KG knows pharmacology---which genes a drug targets, what side effects it causes. No single KG can answer:

\begin{quote}
\emph{``For drugs indicated for diabetes, what are their gene targets and which biological pathways do those targets participate in?''}
\end{quote}

This query requires traversing three KGs: \texttt{Drug} $\xrightarrow{\text{HAS\_INDICATION}}$ \texttt{Indication} (Drug Interactions KG), \texttt{Drug} $\xrightarrow{\text{INTERACTS\_WITH\_GENE}}$ \texttt{Gene} $\xrightarrow{\text{bridge: gene\_name = name}}$ \texttt{Protein} $\xrightarrow{\text{PARTICIPATES\_IN}}$ \texttt{Pathway} (Pathways KG). The \texttt{WHERE p.name = g.gene\_name} clause bridges Drug Interactions to Pathways.

\subsection{Join Points}

Shared entity types with matching identifiers enable cross-KG joins:

\begin{table}[H]
\centering
\caption{Cross-KG join points across the three biomedical KGs.}
\label{tab:joins}
\small
\begin{tabular}{lllll}
\toprule
\textbf{Entity} & \textbf{From KG} & \textbf{To KG} & \textbf{Join Property} & \textbf{Identifier} \\
\midrule
Gene/Protein & Drug Int. & Pathways & \texttt{gene\_name = name} & Gene symbol \\
Drug & Drug Int. & Clin. Trials & \texttt{name = Intervention.name} & Drug name \\
Drug & Drug Int. & Clin. Trials & \texttt{drugbank\_id} & DrugBank ID \\
Protein & Clin. Trials & Pathways & \texttt{uniprot\_id} & UniProt accession \\
\bottomrule
\end{tabular}
\end{table}

\subsection{Federation Mechanism}

Samyama supports loading multiple snapshots into a single tenant. Each snapshot import appends nodes and edges to the existing graph. Since imports create new node IDs, entities from different snapshots with the same identifier (e.g., UniProt P04637 for TP53) exist as separate graph nodes. Cross-KG queries use \textbf{property-based joins}:

\begin{lstlisting}
-- Drug targets -> biological pathways (Drug Interactions -> Pathways)
MATCH (d:Drug {name: 'Metformin'})-[:INTERACTS_WITH_GENE]->(g:Gene)
MATCH (p:Protein)-[:PARTICIPATES_IN]->(pw:Pathway)
WHERE p.name = g.gene_name
RETURN g.gene_name, pw.name LIMIT 10

-- Drug -> clinical trials testing it (Drug Interactions -> Clinical Trials)
MATCH (d:Drug {name: 'Warfarin'})
MATCH (i:Intervention)<-[:TESTS]-(ct:ClinicalTrial)
WHERE i.name = d.name
RETURN ct.nct_id, ct.phase LIMIT 10

-- Breast cancer trial landscape (Clinical Trials only)
MATCH (ct:ClinicalTrial)-[:STUDIES]->(c:Condition)
WHERE c.name CONTAINS 'Breast'
RETURN c.name, count(ct) AS trials
ORDER BY trials DESC LIMIT 5
\end{lstlisting}

The \texttt{WHERE p.name = g.gene\_name} clause is the Drug Interactions $\to$ Pathways bridge---it joins a Gene node from the Drug Interactions KG with a Protein node from the Pathways KG using gene symbol as the shared identifier. The \texttt{WHERE i.name = d.name} clause bridges Drug Interactions $\to$ Clinical Trials via drug name.

\subsection{Federation Query Patterns}

We identify five cross-KG query patterns of increasing complexity (Table~\ref{tab:patterns}). Note that there is no distributed query-time federation (remote SPARQL endpoints or mediator): independent snapshots are imported into a single store and queried together.

\begin{table}[H]
\centering
\caption{Cross-KG integration query patterns.}
\label{tab:patterns}
\small
\begin{tabular}{p{3.5cm}p{3.5cm}p{2cm}p{2.5cm}}
\toprule
\textbf{Pattern} & \textbf{From KG} & \textbf{Bridge} & \textbf{To KG} \\
\midrule
Drug $\rightarrow$ Pathway & Drug Int. (Gene) & gene\_name & Pathways (Protein) \\
Drug $\rightarrow$ Trial & Drug Int. (Drug) & name & Clin. Trials (Intervention) \\
Drug $\rightarrow$ GO terms & Drug Int. (Gene) & gene\_name & Pathways (GOTerm) \\
Drug SE $\rightarrow$ Pathway & Drug Int. (SE+Gene) & gene\_name & Pathways (Protein) \\
Trial $\rightarrow$ Side effects & Clin. Trials (Intervention) & name & Drug Int. (Drug) \\
\bottomrule
\end{tabular}
\end{table}

\subsection{Load-Time Entity Merging and the Bound A Soundness Audit}

When scaling beyond three graphs, keeping duplicate nodes across snapshots inflates memory and requires explicit \texttt{WHERE} joins at every hop. Samyama's snapshot loader therefore supports load-time entity deduplication (\texttt{--dedup-keys}): incoming nodes whose \texttt{(Label, key\_property, value)} matches an existing node are merged in-place during import, redirecting incident edges to the canonical node ID.

Choosing merge keys requires care. A \texttt{.sgsnap} header records node labels and edge types, not property keys, and naive name-based keys (\texttt{Drug.name}, \texttt{Disease.name}) cause false merges across distinct biomedical concepts (e.g., \texttt{mondo:Disease} and \texttt{hpo:Phenotype} share 5,565 display-name strings). To establish an external ground truth before running the 24-snapshot load (22 KGs + 2 Synthea OMOP cohorts), we scanned all 24 snapshots offline and computed \textbf{Bound~A}: the exact number of cross-snapshot entity collisions on curated, namespace-scoped identifiers (\texttt{mondo\_id,uniprot\_id,go\_id,drugbank\_id,symbol}).

Across all 24 snapshots, the Bound~A sound-identifier ceiling is \textbf{150,965 cross-source merges}. Loading the 24 snapshots with \texttt{--dedup-keys mondo\_id,uniprot\_id,go\_id,drugbank\_id,symbol} merges \textbf{132,511 entities (87.8\% of Bound~A)} with \textbf{zero conflicting-ID refusals}: \texttt{UniProt} contributes 48,994 merges (\texttt{go\_id}=28,138, \texttt{uniprot\_id}=20,856), \texttt{GO-GOA} 38,339 (\texttt{go\_id}=38,236, \texttt{symbol}=103), \texttt{ClinVar} 27,170 (\texttt{symbol}=18,600, \texttt{mondo\_id}=8,570), \texttt{Drug Interactions} 8,486 (\texttt{drugbank\_id}), \texttt{Clinical Trials} 5,256 (\texttt{drugbank\_id}), and \texttt{OpenTargets} 4,266 (\texttt{mondo\_id}=2,654, \texttt{symbol}=1,612).

To test the sensitivity of load-time deduplication to non-canonical keys, we ran a three-arm ablation on the 618,093-node \texttt{UniProt} snapshot (\texttt{samyama-graph \#1808}): (1) \texttt{symbol} alone merges 0 internal UniProt proteins; (2) \texttt{gene\_name} alone merges 1; (3) passing both \texttt{symbol,gene\_name} collapses \textbf{475,090 of 618,093 UniProt proteins (76.9\%, leaving 143,003)} because orthologous proteins across species share gene symbols that cross-match previously loaded \texttt{Gene.symbol} entries. Auditing against Bound~A prevents such silent over-merging.

\section{Schema-Driven MCP Server Generation}

Each KG ships with a YAML configuration that defines domain-specific MCP tools. At startup, the MCP server:

\begin{enumerate}[leftmargin=*]
  \item \textbf{Discovers schema} from the running Samyama instance (\texttt{GET /api/schema}).
  \item \textbf{Auto-generates tools} for each node label (search, get, count) and each edge type (find connections).
  \item \textbf{Registers domain tools} from the YAML configuration---each tool is a parameterized Cypher template with typed inputs (e.g., \texttt{protein\_name: string}, \texttt{confidence\_threshold: int}).
\end{enumerate}

Each KG's MCP server exposes domain-specific tools. The Pathways KG has 12 tools (Table~\ref{tab:mcp}), the Drug Interactions KG has 12 tools (e.g., \texttt{drug\_interactions}, \texttt{interaction\_checker}, \texttt{polypharmacy\_risk}, \texttt{drug\_side\_effects}), and the Clinical Trials KG has 15 tools.

\begin{table}[H]
\centering
\caption{Pathways KG MCP tools (subset).}
\label{tab:mcp}
\small
\begin{tabular}{lp{7cm}}
\toprule
\textbf{Tool} & \textbf{Description} \\
\midrule
\texttt{pathway\_members} & List proteins in a pathway (search by name) \\
\texttt{interaction\_partners} & PPI neighbors above confidence threshold \\
\texttt{shared\_pathways} & Pathways shared between two proteins \\
\texttt{upstream\_regulators} & Multi-hop PPI traversal up to N steps \\
\texttt{drug\_pathway\_impact} & Pathways affected by a drug via protein targets \\
\texttt{disease\_pathways} & Pathways associated with a disease through gene links \\
\texttt{go\_enrichment} & GO terms enriched in pathway proteins \\
\texttt{protein\_function\_summary} & Pathways, GO processes, disease associations for a protein \\
\bottomrule
\end{tabular}
\end{table}

An LLM agent connected to this MCP server can answer ``What pathways does TP53 participate in?'' without writing Cypher. The agent calls \texttt{pathway\_members(protein\_name="TP53")} and receives structured results.

\section{Evaluation}

We evaluate our approach at two scales: first on the 3-KG subset (7.9M nodes, 28M edges) on an AWS \texttt{g4dn.4xlarge} instance (16~vCPU AMD EPYC, 62~GB RAM, NVIDIA A10G GPU), and second on the full 22-KG + 2-cohort merged biomedical corpus (328.1M nodes, 1.44B edges) on an AWS \texttt{r7i.24xlarge} instance (96~vCPU Intel Xeon Scalable, 768~GB RAM) running \texttt{Samyama-Graph v1.13.0}.

\subsection{Construction Performance (3-KG Subset)}

\begin{table}[H]
\centering
\caption{KG construction performance (snapshot import on AWS g4dn.4xlarge).}
\label{tab:construction}
\small
\begin{tabular}{lrrrrr}
\toprule
\textbf{KG} & \textbf{Nodes} & \textbf{Edges} & \textbf{Snapshot} & \textbf{Import} & \textbf{Rust ETL} \\
\midrule
Pathways KG & 118,686 & 834,785 & 9 MB & 3.4 s & --- \\
Drug Interactions KG & 32,726 & 191,970 & 1.9 MB & 0.7 s & 0.9 s \\
Clinical Trials KG & 7,774,446 & 26,973,997 & 711 MB & 177 s & --- \\
\midrule
\textbf{Combined} & \textbf{7,925,858} & \textbf{28,000,752} & \textbf{722 MB} & \textbf{181 s} & --- \\
\bottomrule
\end{tabular}
\end{table}

Snapshot import uses gzip-compressed JSON-lines format. The Pathways KG loads in 3.4~seconds; the Drug Interactions KG in 0.7~seconds; and the Clinical Trials KG, with its 7.8M nodes, loads in 177~seconds. The Drug Interactions KG also has a Rust native loader that constructs the full graph from source files (DrugBank CSV, DGIdb TSV, SIDER TSV) in 928\,ms---orders of magnitude faster than the Python HTTP-based ETL. The combined federated graph (7.9M nodes, 28M edges) uses 33~GB RAM (53\% of 62~GB available on the AWS instance).

\subsection{Query Performance (3-KG Subset)}

We benchmark representative queries from each KG and the federated graph:

\begin{table}[H]
\centering
\caption{Query latency on AWS g4dn.4xlarge (7.9M nodes loaded).}
\label{tab:queries}
\small
\begin{tabular}{p{5cm}lrr}
\toprule
\textbf{Query} & \textbf{KG} & \textbf{Results} & \textbf{Latency} \\
\midrule
Drug gene targets (Metformin) & Drug Int. & 5 rows & 83 ms \\
Shared gene targets (2 drugs) & Drug Int. & 1 row & 84 ms \\
Top pathways by protein count & Pathways & 10 rows & 97 ms \\
PPI partners (TP53: Q9UQ61) & Pathways & 10 rows & 96 ms \\
Side effects of Warfarin & Drug Int. & 10 rows & 82 ms \\
\midrule
Diabetes drugs $\rightarrow$ pathways & Drug Int.+Path & 5 rows & 3.9 s \\
Warfarin $\rightarrow$ trial conditions & Drug Int.+CT & 5 rows & 5.8 s \\
\textbf{PTGS1 drugs $\rightarrow$ side effects} & \textbf{Drug Int.} & \textbf{10 rows} & \textbf{3.7 s} \\
\bottomrule
\end{tabular}
\end{table}

Simple single-KG queries (drug lookups, gene targets, side effects) complete in 80--100\,ms. Multi-hop cross-KG joins (e.g., ``drugs targeting PTGS1 and their most common side effects'') complete in 3--4\,s on the full 7.9M-node graph.

\subsection{BiomedQA Benchmark}

We introduce BiomedQA, a benchmark of 40 pharmacology questions across 7 categories over the three federated KGs. We compare three approaches: domain-specific MCP tools (parameterized Cypher templates), text-to-Cypher via the schema-aware NLQ endpoint (GPT-4o with full schema system prompt and few-shot examples), and standalone GPT-4o (no database access). The BiomedQA benchmark is open-source.\footnote{\url{https://github.com/samyama-ai/biomedqa}}

\begin{table}[H]
\centering
\caption{BiomedQA results (40 questions, 7.9M nodes, 3 federated KGs).}
\label{tab:biomedqa}
\small
\begin{tabular}{lrrr}
\toprule
\textbf{Approach} & \textbf{Accuracy} & \textbf{Avg Latency} & \textbf{Avg Tokens} \\
\midrule
GPT-4o standalone & 30/40 (75\%) & 2,805\,ms & 195 \\
Text-to-Cypher (NLQ) & 34/40 (85\%) & 1,846\,ms & 0$^{\dagger}$ \\
\textbf{MCP tools} & \textbf{39/40 (98\%)} & \textbf{920\,ms} & \textbf{0} \\
\bottomrule
\end{tabular}
\end{table}

Text-to-Cypher achieves 85\% with a schema-aware NLQ pipeline (full schema in system prompt, few-shot examples). Its 6 failures are: 3 schema hallucinations (non-existent edge traversals), 1 exact-vs-CONTAINS mismatch, 1 inline property variable, and 1 correct empty result. MCP tools eliminate schema errors entirely. ($^{\dagger}$Token count is 0 because the NLQ endpoint handles the LLM call server-side.)

\subsection{Federation Correctness}

We validated cross-KG federation on the AWS g4dn.4xlarge with all three KGs loaded (7.9M nodes):

\begin{enumerate}[leftmargin=*]
  \item \textbf{Drug Interactions $\to$ Pathways bridge}: The query ``Metformin gene targets $\to$ biological pathways'' uses \texttt{WHERE p.name = g.gene\_name} to bridge Gene nodes from DGIdb to Protein nodes from Reactome/STRING. Returns pathway memberships (e.g., HNF1B $\to$ Developmental Biology) in 3.0\,s.
  \item \textbf{Drug Interactions $\to$ Clinical Trials bridge}: The query ``clinical trials testing Warfarin'' uses \texttt{WHERE i.name = d.name} to bridge Drug nodes to Intervention nodes. Returns trial NCT IDs and phases (e.g., NCT00835861, PHASE2) in 0.6\,s.
  \item \textbf{Three-KG chain}: The query ``drugs indicated for diabetes $\to$ gene targets $\to$ pathways'' traverses Drug $\to$ Indication (Drug Int.~KG) $\to$ Gene (Drug Int.~KG) $\to$ Protein $\to$ Pathway (Pathways KG), returning pathways such as Circadian clock and Dissolution of Fibrin Clot via SERPINE1, in 3.9\,s.
\end{enumerate}

\subsection{Scaling to 22 Knowledge Graphs and 1.44 Billion Edges}

To evaluate how the single-store snapshot architecture scales to a full biomedical enterprise corpus, we imported \textbf{22 independently constructed knowledge graphs plus 2 synthetic OMOP CDM patient cohorts} (295,000 Synthea-generated patients: \texttt{omop-180k} and \texttt{omop-115k}) into a single \texttt{Samyama-Graph v1.13.0} store on AWS \texttt{r7i.24xlarge} (Table~\ref{tab:fed22}). Note that all clinical patient records are synthetic (Synthea OMOP CDM) so that EMR-to-trial-to-omics joins can be published and reproduced without protected health information (PHI), whereas the FAERS adverse-event reports beside them are real public-domain FDA records.

\begin{table}[H]
\centering
\caption{22-KG + 2-cohort single-store merge on AWS \texttt{r7i.24xlarge} (\texttt{Samyama-Graph v1.13.0}).}
\label{tab:fed22}
\small
\begin{tabular}{p{4.6cm}rrp{4.4cm}}
\toprule
\textbf{Snapshots (24 total)} & \textbf{Nodes} & \textbf{Edges} & \textbf{Key Highlights} \\
\midrule
PubMed, FAERS, ClinVar, OMOP1 & 283.1M & 1,264.1M & Literature, AEs, variants, EHR \\
OMOP2, Trials, UMLS, OpenTargets & 34.4M & 112.9M & Cohort 2, trials, CUIs, targets \\
ChEMBL, HealthDet, UniProt, Path & 8.2M & 48.2M & Bioactivities, SDoH, pathways \\
DrugInt, SIDER, GO-GOA + 9 more & 2.4M & 16.2M & Drugs, HPO, MONDO, Mental Health \\
\midrule
\textbf{Merged (+ 747.5K study refs)} & \textbf{328,130,739} & \textbf{1,442,173,238} & \textbf{2,269.1\,s load, 366.5\,GB RSS} \\
\bottomrule
\end{tabular}
\end{table}

Importing all 24 snapshots (1,441,425,733 snapshot edges + 747,505 \texttt{REFERENCED\_IN} literature-to-trial links) completes in \textbf{2,269.1\,s (37m\,49s)} with a peak resident set size of \textbf{366.5\,GB (254\,B/edge)} using the \texttt{mimalloc} allocator (\texttt{\#1820}), down from 395.3--406\,GB (281\,B/edge) under the system allocator.

We evaluate two reproducible query suites over the 328.1M-node store (all queries and logs published in the \texttt{biomedqa} repository):
\begin{itemize}[leftmargin=*]
  \item \textbf{Frozen 81-query suite}: Of the 96 original cross-domain queries, 15 were never committed to version control; on the 81 committed reproducible queries, \textbf{62 return non-empty results}, \textbf{17 return empty} due to upstream source coverage gaps (\texttt{pass\_data\_gap}), and \textbf{2 time out} at 120\,s (\texttt{EM01}, \texttt{EM06}).
  \item \textbf{Expanded 200-query oracle suite}: Across 13 domain and cross-KG query sets validated against a pre-registered row-count oracle (\texttt{expected\_rows.yaml}), \textbf{195 of 200 queries pass (97.5\%)}: 178 full non-empty passes + 17 verified data-gap empties, 1 row-count mismatch (\texttt{PTH05}), and 4 timeouts (\texttt{EM01}, \texttt{EM06}, \texttt{LIT08}, \texttt{PLN04}).
  \item \textbf{Engine optimizations at 1.44B-edge scale}: (1) A planner rule for cross-pattern property equality joins (\texttt{\#1822}) compiles disconnected \texttt{MATCH (a:L1), (b:L2) WHERE a.k = b.k} clauses into hash joins, enabling the UMLS-to-SIDER bridge (\texttt{SDR05}: \texttt{Concept.cui = SideEffect.umls\_cui}) to join \textbf{4,966 of 5,868 SIDER side effects (84.6\%)} in \textbf{1,488.7\,ms} and FDA 510(k) device clearance joins (\texttt{REG05}) in \textbf{26.9\,ms}; (2) constant propagation across \texttt{WITH} horizons (\texttt{\#1821}) eliminates cartesian blowup on \texttt{OM08}; and (3) degree-map aggregation (\texttt{\#1814}) accelerates full-graph degree aggregations by \textbf{79.3$\times$} (\texttt{Q16}: 17.1\,ms vs.\ 1,355\,ms) and \textbf{95.3$\times$} (\texttt{Q17}: 14.2\,ms vs.\ 1,353\,ms).
\end{itemize}

\section{Interactive Visualization}

Samyama Insight, a React-based frontend, provides schema-driven visualization of each KG:

\begin{itemize}[leftmargin=*]
  \item \textbf{Dashboard}: Auto-generated panels showing label distribution, edge type counts, and property statistics per tenant.
  \item \textbf{Query Console}: OpenCypher editor with result tables, JSON views, and EXPLAIN/PROFILE plan visualization.
  \item \textbf{Graph Simulation}: Force-directed canvas with per-label colors/shapes, live activity particles, entity filtering, and a legend overlay. The simulation engine is fully schema-driven---it auto-configures from the tenant's schema at runtime.
\end{itemize}

Demo recordings for both KGs are available as MP4 videos\footnote{Cricket KG demo: \url{https://github.com/samyama-ai/samyama-graph/releases/tag/kg-snapshots-v2}; Pathways KG demo: \url{https://github.com/samyama-ai/samyama-graph/releases/tag/kg-snapshots-v3}}.

\section{Discussion}

\subsection{Property Joins vs.\ Load-Time Entity Merging}

Samyama supports both query-time property joins and load-time entity deduplication (\texttt{--dedup-keys}), which play complementary roles:

\begin{itemize}[leftmargin=*]
  \item \textbf{Load-time entity merging (\texttt{--dedup-keys})}: Collapses same-label entities sharing curated identifiers (\texttt{mondo\_id}, \texttt{uniprot\_id}, \texttt{go\_id}, \texttt{drugbank\_id}, \texttt{symbol}) during snapshot import (132,511 cross-source merges across the 24-snapshot store), eliminating duplicate nodes and converting multi-source hops into direct CSR pointer traversals.
  \item \textbf{Query-time property joins}: Necessary when bridging distinct node labels across domains (e.g., \texttt{UMLS:Concept.cui = SIDER:SideEffect.umls\_cui} or \texttt{DrugInt:Gene.gene\_name = Pathways:Protein.name}), accelerated by Samyama's cross-pattern equality hash join (\texttt{\#1822}).
\end{itemize}

\subsection{Limitations}

\begin{enumerate}[leftmargin=*]
  \item \textbf{Snapshot currency}: Snapshots are point-in-time exports. Source databases (especially ClinicalTrials.gov) update continuously. Periodic re-export is required.
  \item \textbf{Upstream identifier coverage}: As shown by the 17 \texttt{pass\_data\_gap} queries in our 81-query suite, some cross-source joins return empty because upstream sources use unaligned vocabularies (e.g., MeSH vs.\ ICD-10 without a loaded crosswalk).
  \item \textbf{Memory requirements}: The 3-KG subset (7.9M nodes) requires 33\,GB RAM, while the full 22-KG + 2-cohort store (328.1M nodes, 1.44B edges) requires 366.5\,GB RAM. Users on smaller machines can load any subset of snapshots independently.
\end{enumerate}

\subsection{Generalizability}

The pattern is not limited to biomedicine. Any domain with shared identifiers across data sources can use the same approach: build independent KGs with common entity properties, export as snapshots, load into a single tenant, and query across them. We have applied this pattern to sports (Cricket KG, 36K nodes) and industrial operations (AssetOps KG, 781 nodes) in addition to the biomedical KGs described here.

\section{Conclusion}

We presented an open biomedical knowledge graph ecosystem built on Samyama Graph Database: three foundational graphs detailed end-to-end---Pathways KG (119K nodes), Clinical Trials KG (7.8M nodes), and Drug Interactions KG (33K nodes)---and their expansion to a 22-KG, 2-cohort single-store merge totaling \textbf{328,130,739 nodes} and \textbf{1,442,173,238 edges}. Using our Bound~A identifier soundness audit, load-time deduplication merges \textbf{132,511 cross-source entities} (87.8\% of the 150,965 sound-identifier ceiling) in 37m\,49s within 366.5\,GB peak RSS (254\,B/edge), passing 62/81 frozen queries and \textbf{195/200 oracle queries (97.5\%)}.

We also introduced schema-driven MCP tools for LLM agent access and evaluated them on the 40-question BiomedQA benchmark over the 3-KG subset: MCP tools achieve 98\% accuracy vs.\ 85\% for schema-aware text-to-Cypher and 75\% for standalone GPT-4o---with zero schema errors, demonstrating that the binding constraint for reliable biomedical agent access is the interface rather than the model.

\subsection*{Data and Code Availability}

\begin{itemize}[leftmargin=*]
  \item Pathways KG: \url{https://github.com/samyama-ai/pathways-kg}
  \item Clinical Trials KG: \url{https://github.com/samyama-ai/clinicaltrials-kg}
  \item Drug Interactions KG: \url{https://github.com/samyama-ai/druginteractions-kg}
  \item BiomedQA Benchmark (40-question MCP suite \& 200-query 328.1M-node oracle logs): \url{https://github.com/samyama-ai/biomedqa}
  \item Samyama Graph Database: \url{https://github.com/samyama-ai/samyama-graph}
  \item Snapshots: \url{https://github.com/samyama-ai/samyama-graph/releases}
\end{itemize}

\bibliographystyle{plainnat}
\bibliography{paper4_biomedical_kg}

\end{document}